\documentclass[smallextended,final]{svjour3}
\let\makeheadbox\relax
\smartqed
\usepackage{amsmath, amssymb}
\usepackage{bm}
\usepackage[T1]{fontenc}               
\usepackage[utf8]{inputenc}            
   
\usepackage{paralist}
\usepackage{setspace}
   \setdisplayskipstretch{2}
\usepackage[dvipsnames]{xcolor}        
\usepackage{Z-fonts}
\usepackage[final,%
            colorlinks=true,%
            linkcolor=RedViolet,%
            citecolor=RedViolet,
            urlcolor=RoyalBlue]{hyperref}   

  \DeclareMathSymbol{A}{\mathalpha}{operators}{`A}
  \DeclareMathSymbol{B}{\mathalpha}{operators}{`B}
  \DeclareMathSymbol{C}{\mathalpha}{operators}{`C}
  \DeclareMathSymbol{D}{\mathalpha}{operators}{`D}
  \DeclareMathSymbol{E}{\mathalpha}{operators}{`E}
  \DeclareMathSymbol{F}{\mathalpha}{operators}{`F}
  \DeclareMathSymbol{G}{\mathalpha}{operators}{`G}
  \DeclareMathSymbol{H}{\mathalpha}{operators}{`H}
  \DeclareMathSymbol{I}{\mathalpha}{operators}{`I}
  \DeclareMathSymbol{J}{\mathalpha}{operators}{`J}
  \DeclareMathSymbol{K}{\mathalpha}{operators}{`K}
  \DeclareMathSymbol{L}{\mathalpha}{operators}{`L}
  \DeclareMathSymbol{M}{\mathalpha}{operators}{`M}
  \DeclareMathSymbol{N}{\mathalpha}{operators}{`N}
  \DeclareMathSymbol{O}{\mathalpha}{operators}{`O}
  \DeclareMathSymbol{P}{\mathalpha}{operators}{`P}
  \DeclareMathSymbol{Q}{\mathalpha}{operators}{`Q}
  \DeclareMathSymbol{R}{\mathalpha}{operators}{`R}
  \DeclareMathSymbol{S}{\mathalpha}{operators}{`S}
  \DeclareMathSymbol{T}{\mathalpha}{operators}{`T}
  \DeclareMathSymbol{U}{\mathalpha}{operators}{`U}
  \DeclareMathSymbol{V}{\mathalpha}{operators}{`V}
  \DeclareMathSymbol{W}{\mathalpha}{operators}{`W}
  \DeclareMathSymbol{X}{\mathalpha}{operators}{`X}
  \DeclareMathSymbol{Y}{\mathalpha}{operators}{`Y}
  \DeclareMathSymbol{Z}{\mathalpha}{operators}{`Z}

  \newcommand\GL[1]{{\mathrm{GL}_{#1}(\RR)}} 
\renewcommand\O[1]{{\mathrm{O}_{#1}}}        
  \newcommand\RR{{\mathbf R}}                
  \newcommand\SO[1]{{\mathrm{SO}_{#1}}}      
  \newcommand\ZZ{{\mathbf Z}}                
  
  \newcommand\Lie{\mathfrak}
  \newcommand\LG{{\Lie{g}}}
  \newcommand\LGL[1]{{\Lie{gl}_{#1}(\RR)}}
  \newcommand\LK{{\Lie{k}}}
  \newcommand\LM{{M}}
  \newcommand\LO[1]{{\Lie{o}_{#1}}}
  \newcommand\LP{{\Lie{p}}}

\DeclareMathOperator\Ad{Ad}
\DeclareMathOperator\Tr{Trace}
  \newcommand\1{{\mathbf1}}                               
  \newcommand\C{{C}}                                      
  \newcommand\e[1]{{\mathrm e^{\hspace{.06em}#1}}}        
  \newcommand\g{{\mathrm g}}                              
  \newcommand\inv{^{-1}}                                  
\renewcommand\l{\lambda}                                  
  \newcommand\s{\sigma}                                   

  \newcommand\<{\langle}                                  
\renewcommand\>{\rangle}                                  
  \newcommand\implique{\text{ implies }}
  
\usepackage{etoolbox}
\makeatletter
\newcommand*{\@linkedbibitem}[1]{%
  \def\this@biblink{#1}%
  \bibitem}
\newcommand*{\linkedbibitem}{\hyper@normalise\@linkedbibitem}
\renewcommand*{\@BIBLABEL}[1]{%
  \ifdefvoid\this@biblink
    {[#1]}
    {[\expandafter\href\expandafter{\this@biblink}%
       {#1}]}}
\makeatother

\begin{document}
\title{Relativity without light: A new proof of Ignatowski's theorem}
\author{Jean-Philippe Anker \and François Ziegler}
\institute{J.-Ph. Anker \at
           Institut Denis Poisson,
           Université d'Orléans,
           B.P. 6759,
           45067 Orléans cedex 2, France \\
           \email{anker@univ-orleans.fr}
           \and
           F. Ziegler \at
           Department of Mathematical Sciences,
           Georgia Southern University,
           Statesboro, GA 30460, U.S.A. \\
           \email fziegler@georgiasouthern.edu
}
\date{August 7, 2020}

\maketitle
\begin{abstract}
   V.~Ignatowski (1910) showed that assumptions about light are not necessary to obtain Lorentzian kinematics as one of only few possibilities. We give a much simplified proof of his result as formulated by V.~Gorini (1971) for $n$+1-dimensional space-time.
\end{abstract}

\section{Introduction}
\setcounter{equation}{2}

The Lorentz group of space-time transformations emerged progressively in work of Voigt, Larmor and Lorentz on the symmetry of Maxwell's equations. This was subtle business, as the geometrical nature of the electromagnetic field (a $2$-form) had yet to be elucidated, so as Lorentz recalls in \cite[p.\,297]{Lorentz:1921}: 
\begin{quote}
   {\small For other physical quantities such as electric and magnetic forces, a less direct method must be followed; one will seek, perhaps a little by trial and error, the transformation formulas suitable for ensuring the invariance of the electromagnetic equations.}
\end{quote}
As one knows, a drastic simplification occurred when Einstein, Poincaré and Minkowski characterized the group as those transformations which
\begin{compactenum}[(1)]
   \setcounter{equation}{1}
   \item[(\theequation)] \label{affine} are affine (so they take the straight world-lines of free particles to other straight world-lines, respecting the law of inertial motion);\pagebreak
   \refstepcounter{equation}
   \item[(\theequation)] preserve the cones
   \begin{equation}
      \label{metric}
      0 = dt^2 - \s\|d\bm r\|^2,
      \quad\text{where}\quad
      \s = c^{-2}
   \end{equation}
   (so they take world-lines with speed $c$ to other such world-lines, respecting the law of light propagation).
\end{compactenum}
Soon after, V.~Ignatowski \cite{Ignatowski:1910,Ignatowski:1911} added the remarkable observation that the existence of a (possibly infinite) invariant speed --- leading to the Lorentz and Galilei groups as essentially the only possibilities --- is in fact a consequence of \eqref{affine} and symmetry under Euclidean displacements alone, regardless of any considerations involving the propagation of light.

While this is conceptually comforting (the question whether light actually travels at the invariant speed is an experimental one \cite{Goldhaber:2010}), a drawback of Ignatowski's original argument is that it was essentially 1+1-dimen\-sio\-nal. As such it has the distinction of being one of the most often rediscovered in mathematical physics,\footnote{Despite duly appearing in the standard references \cite[§4]{Pauli:1921}, \cite[p.\,43]{Whittaker:1953}, \cite[p.\,206]{Miller:1981}.} but a clear-cut version valid in 3+1-dimensional space-time had to wait until V.~Gorini \cite{Gorini:1971,Gorini:1973} proved in substance the following:

\begin{theorem}
   \label{homogeneous_theorem}
   Suppose $n\geqslant2$ and let $G$ be a subgroup of $\GL{n+1}$ such that
   \begin{equation}
      \label{homogeneous_hypothesis}
   	G\cap\begin{pmatrix}\GL{n}&\,0\\0&\,\RR^\times\end{pmatrix}
   	=\begin{pmatrix}\O{n}&\,\,0\\0&\,\,\pm1\end{pmatrix}.
   \end{equation}
   Write $K$ for the right-hand side of \eqref{homogeneous_hypothesis}. Then either $G=K$ or there is a number $\s\in\RR\cup\{\infty\}$ such that $G=K\exp(\LP_\s)$\textup, where
   \begin{equation}
      \label{LP}
      \LP_\s=\left\{
      \begin{pmatrix}
         0&\,\,\bm b\\\s{}^{\textup t}\bm b&\,\,0
      \end{pmatrix}:\bm b\in\RR^n
      \right\},
      \qquad\quad
      \LP_\infty=\left\{
      \begin{pmatrix}
         0&\,\,0\\{}^{\textup t}\bm c&\,\,0
      \end{pmatrix}:\bm c\in\RR^n
      \right\}.\footnote{Here and elsewhere ${}^{\textup t}m$ always means the transpose of any row, column or matrix $m$; and $\O{n}$ denotes the orthogonal group $\{A\in\GL{n}:{}^{\textup t}AA=\1\}$.}
   \end{equation}   
\end{theorem}
This result seems far less known than it deserves to be --- perhaps because Gorini's proof is too tedious to widely reproduce. It says that the world's kinematical group (by which we mean, any group $G$ satisfying the theorem's hypotheses) must be isomorphic to one of only 5 possibilities:
\begin{enumerate}[\quad(a)]
   \item if $\s>0$, the Lorentz group $\O{n,1}$ as named in \cite{Poincare:1906};
   \item if $\s=0$, the homogeneous Galilei group \cite{Frank:1908};
   \item if $\s<0$, the orthogonal group $\O{n+1}$ \cite{Jordan:1870};
   \item if $\s=\infty$, the homogeneous Carroll group \cite{Levy-Leblond:1965};
   \item if $G=K$, the homogeneous Aristotle group \cite{Souriau:1970}.
\end{enumerate}
Our purpose is to give a simpler proof, which we manage for two main reasons. First, we bring to bear a theorem of Bourbaki \cite{Bourbaki:1972} which endows $G$ with a Lie group structure. This justifies \emph{a posteriori} the use of Lie algebra methods pioneered by V.~Lalan \cite{Lalan:1937}. Secondly, we will see that computations left out by Lalan can be shortened by applying a modicum of representation theory.

Over earlier 3+1-dimensional treatments, Gorini's formulation has the advantage of concision: his hypotheses can all be stated before the proof starts, rather than introduced piecemeal as ``postulates'' along the steps of a long-winded discussion. (E.g.~Hahn \cite{Hahn:1913} has 7 axioms spread over 14 pages.) In~other words, his is a genuine mathematical theorem, and readers so inclined can skip straight to our proof in §2. Nevertheless we feel that some discussion of its hypotheses and their significance is warranted, so we devote the rest of this Introduction to that.

\subsection{The linearity assumption}

The first key assumption of Theorem \ref{homogeneous_theorem} is to consider only linear transformations of space-time $\RR^{n+1}$. This is in fact an oversimplification designed to ease the exposition: as \eqref{affine} suggests, the true setting is affine transformations; i.e.~we should really replace every group $\Gamma\subset\GL{n+1}$ in sight by its \emph{inhomogeneous} avatar, the semidirect product 
\begin{equation}
   \label{semidirect}
   \Gamma\ltimes\RR^{n+1}
   \cong
   \begin{pmatrix}
      \Gamma&\,\,\,\RR^{n+1}\\
      0&\,\,\,1
   \end{pmatrix}
   \subset\GL{n+1}\ltimes\RR^{n+1}
\end{equation}
and prove:
\begin{theorem}
   \label{inhomogeneous_theorem}
   Suppose $n\geqslant2$ and let $G$ be a subgroup of $\GL{n+1}\ltimes\RR^{n+1}$ such that
   \begin{equation}
      \label{inhomogeneous_hypothesis}
   	G\cap\left(
   	\begin{pmatrix}
      	\GL{n}&\,0\\
      	0&\,\RR^\times
   	\end{pmatrix}
   	\ltimes\RR^{n+1}\right)
   	=
   	\begin{pmatrix}
   	   \O{n}&\,\,0\\
   	   0&\,\,\pm1
   	\end{pmatrix}
   	\ltimes\RR^{n+1}.
   \end{equation}
   Define $K$ and $\LP_\s$ as in Theorem \ref{homogeneous_theorem}. Then either $G=K\ltimes\RR^{n+1}$ or there is a number $\s\in\RR\cup\{\infty\}$ such that $G=K\exp(\LP_\s)\ltimes\RR^{n+1}$.
\end{theorem}

This can be deduced from Theorem \ref{homogeneous_theorem} as an easy corollary, or maybe better, proved simultaneously by adding throughout a row and column as in \eqref{semidirect}. The meaning of the assumption is that we are looking for ``symmetries of Newton's first law'' (of uniform rectilinear motion), and its justification is the Fundamental Theorem of Affine Geometry, which says that a transformation of $\RR^{n+1}$ is affine if and only if it maps straight lines to straight lines \cite[Thm 2.6.3]{Berger:1987}. One might object that Newton's first law is only observed at infraluminal speeds, but G.~Hegerfeldt \cite{Hegerfeldt:1972} has shown that a transformation mapping ``slow'' lines to lines necessarily maps all lines to lines.

\subsection{The Euclidean invariance assumption}

The theorem's second key assumption, \eqref{homogeneous_hypothesis} or properly \eqref{inhomogeneous_hypothesis}, is really twofold. It says firstly that $G$ contains the ``Aristotle'' group, i.e.~Euclidean motions and reflections as well as time translations and reversals; this is expected insofar as these are symmetries of known physical laws. Secondly it says that $G$ contains \emph{no other} transformations not mixing space and time. Here one might object that by not allowing independent changes of units in space and time (which would destroy the result) we are \emph{of course} smuggling in an invariant speed.

That would be misunderstanding, however, as we are not after ``all possible changes of variables'', but after \emph{transformations taking a possible system to another possible system}. As one knows (today!) an inflated atom is not a possible atom; these and other (``passive'') changes of description have their place in physics, but not necessarily in a \emph{group} including space-time transformations.

\subsection{The group property}

The last remark points to the subtlety of the theorem's third key assumption: the transformations of interest make a group. Today groups are in the physicists' DNA, and here is not the place for an epistemological discussion of why that should be. (We recommend the one in \cite[pp.\,18--20]{Fell:1988b}.) But to those for whom Ignatowski showed that ``Galileo could have derived special relativity'' one must make the objection of anachronism: in point of fact the word \emph{group} did not enter the picture until the papers \cite{Einstein:1905a,Poincare:1906}, and Galilei transformations themselves were not singled out or named until later \cite{Frank:1908,Minkowski:1909}.

\section{Proof of Theorem \ref{homogeneous_theorem}}

\subsection{Lie group structure of $G$}

The first key fact to be used is that $G$ admits a canonical (``initial'') Lie group structure having Lie algebra
\begin{equation}
   \label{LG}
   \LG=\left\{Z\in\LGL{n+1}:\e{tZ}\in G \text{ for all } t\in\RR\right\}.
\end{equation}
This remarkable theorem of \cite[§III.4.5]{Bourbaki:1972} is exposed again in \cite[§2.2]{Rossmann:2002}, \cite[§9.6.2]{Hilgert:2012a}, \cite[§6.14]{Godement:2017}. We emphasize that it is valid for any subgroup $G$ of any Lie group, not \emph{a priori} closed nor endowed with the subspace topology.

\subsection{Determination of the Lie algebra $\LG$}

By \eqref{homogeneous_hypothesis} $\LG$ contains the Lie algebra $\LK$ of $K$. We claim that either $\LG = \LK$~or
\begin{equation}
   \label{k_plus_p}
   \LG = \LK \oplus \LP_\s
\end{equation}
for some $\s\in\RR\cup\{\infty\}$. Indeed, deriving $k\e{tZ}k\inv$ at $t=0$ shows that \eqref{LG} is an invariant subspace of $\LGL{n+1}$ for the adjoint representation of $\O{n}\subset K$:
\begin{equation}
   \label{Ad}
   \Ad
   \begin{pmatrix}R&\,\,0\\0&\,\,1\end{pmatrix}
   \begin{pmatrix}A&\,\,\bm b\\{}^{\textup t}\bm c&\,\,d\end{pmatrix}
   =
   \begin{pmatrix}RAR\inv&\,\,R\bm b\\{}^{\textup t}(R\bm c)&\,\,d\end{pmatrix},
   \rlap{\quad$R\in\O{n}$.}
\end{equation}
Therefore we have $\LG=\bigoplus_{i=0}^3(\LM_i\cap\LG)$ where $\LGL{n+1}=\bigoplus_{i=0}^3 \LM_i$ is the decomposition of $\LGL{n+1}$ into isotypic components (i.e.~multiples of irreducibles) under $\O{n} $\cite[§4, Prop.~4d]{Bourbaki:2012}. Here the summands are, when $n\geqslant2$,  
\begin{subequations}
	\begin{align}
	   \LM_0 &=
	   \left\{\begin{pmatrix}
	   \lambda\1&\,\,0\\0&\,\,\mu
	   \end{pmatrix}:\lambda, \mu\in\RR\right\}
	   \\
	   \LM_1 &=
	   \left\{\begin{pmatrix}
	   A&\,\,0\\0&\,\,0
	   \end{pmatrix}:A \text{ skew-symmetric}\right\}=\LK
	   \\
	   \LM_2 &=
	   \left\{\begin{pmatrix}
	   A&\,\,0\\0&\,\,0
	   \end{pmatrix}:A \text{ symmetric}, \Tr(A)=0\right\}
	   \\
	   \LM_3 &=
	   \left\{\begin{pmatrix}
	   0&\,\,\bm b\\{}^{\textup t}\bm c&\,\,0
	   \end{pmatrix}:\bm b, \bm c\in\RR^n\right\}
	\end{align}
\end{subequations}
(see e.g.~\cite[Prop.~1.105]{Besse:1987}). Now clearly $\LM_1\subset\LG$, and $\LM_0\cap\LG=\LM_2\cap\LG=\{0\}$: if $\LG$ contained any nonzero members of $\LM_0$ or $\LM_2$ then $G$ would contain their exponentials, which is excluded by  \eqref{homogeneous_hypothesis}. Next we claim that any $Z\in \LM_3\cap\LG$ has $\bm b$ and $\bm c$ collinear. To see this, put $A=\bm b{}^{\textup t}\bm c-\bm c{}^{\textup t}\bm b\in\LO{n}$ and compute
\begin{equation}
   \label{collinearity}
   \Bigl[\begin{pmatrix}
   0&\,\,\bm b\\{}^{\textup t}\bm c&\,\,0
   \end{pmatrix},\Bigl[\begin{pmatrix}
   0&\,\,\bm b\\{}^{\textup t}\bm c&\,\,0
   \end{pmatrix},\begin{pmatrix}
   A&\,\,0\\0&\,\,0
   \end{pmatrix}\Bigr]\Bigr]=\begin{pmatrix}
   *&\,\,0\\0&\,\,2\left\{\|\bm b\|^2\|\bm c\|^2-({}^{\textup t}\bm b\bm c)^2\right\}
   \end{pmatrix}.
\end{equation}
As this is contained in $[\LG,[\LG,\LK]]\subset\LG$, the lower right entry must be $0$: so the Cauchy-Schwarz bound is attained, i.e.~$\bm b$ and $\bm c$ are indeed collinear. Thus each $Z\in \LM_3\cap\LG$ is in $\LP_\s$ for some $\s$, which we claim must be the same for any two nonzero members $Z_1, Z_2$: else, considering linear combinations of $Z_1$ and $\Ad(k)(Z_2)$ \eqref{Ad} would readily show that $\LG$ contains all of $\LM_3$ and hence equals $\LM_1\oplus \LM_3$, which by \eqref{collinearity} is not a Lie subalgebra. So \eqref{k_plus_p} is proved.

\subsection{End of proof when $\s\in\{0,\infty\}$}

The key technique we use to obtain $G$ from $\LG$ is that (as one sees by deriving $g\e{tZ}g\inv$ at $t=0$) $G$ must always be contained in the \emph{normalizer}
\begin{equation}
   \label{normalizer}
   N(\LG)=\left\{a\in\GL{n+1}:a\LG a\inv\subset\LG\right\}.
\end{equation}
Assume $\s=0$ and
$a=\left(
\begin{smallmatrix}
   U&\bm v\\
   {}^{\textup t}\bm w&x
\end{smallmatrix}
\right)\in N(\LG)$. This means that for every
$Z=\left(
\begin{smallmatrix}
   A&\bm b\\
   0&0
\end{smallmatrix}
\right)\in\LG$ there is $Z'\in\LG$ such that $aZ=Z'a$, i.e.
\begin{equation}
   \begin{pmatrix}
      UA&\,\,U\bm b\\
      {}^{\textup t}\bm wA&\,\,{}^{\textup t}\bm w\bm b
   \end{pmatrix}=
   \begin{pmatrix}
      A'U+\bm b'{}^{\textup t}\bm w&\,\,A'\bm v+\bm b'x\\
      0&\,\,0
   \end{pmatrix}.
\end{equation}
Therefore $\bm w=0$ and so every member of $G$ writes 
$\left(
\begin{smallmatrix}
   U&\bm v\\
   0&x
\end{smallmatrix}
\right)=
\left(
\begin{smallmatrix}
   U&0\\
   0&x
\end{smallmatrix}
\right)\exp\left(
\begin{smallmatrix}
   0&U\inv\bm v\\
   0&0
\end{smallmatrix}
\right)
$
as required: since \eqref{LG} ensures the second factor is in $G$, so must the first which is therefore in
$\left(
\begin{smallmatrix}
   \O{n}&0\\
   0&\pm1
\end{smallmatrix}
\right)
$
by hypothesis \eqref{homogeneous_hypothesis}. The case $\s=\infty$ is similar.

\subsection{Computation of the normalizer $N(\LG)$ when $\s\in\RR\smallsetminus\{0\}$}

To facilitate this computation, let us introduce on $\RR^{n+1}$ the two inner products $\<x,y\>_\pm ={}^{\textup t}x\g_\pm y$ where
\begin{equation}
   \label{g}
   \g_\pm = \begin{pmatrix}
   \mp\s\1&\,\,0\\0&\,\,1
   \end{pmatrix}
\end{equation}
and write $Z^\pm$ for the resulting adjoints of $Z\in\LGL{n+1}$, defined by the relation $\<x,Zy\>_\pm=\<Z^\pm x,y\>_\pm$ or more explicitly
\begin{equation}
	Z^\pm=\g_\pm\inv{}^{\textup t}Z\g_\pm^{\phantom\pm},
	\qquad\quad
	\begin{pmatrix}A&\,\,\bm b\\{}^{\textup t}\bm c&\,\,d\end{pmatrix}^\pm=
   \begin{pmatrix}\smash{{}^{\textup t}A}&\,\,\mp\bm c/\s\\\mp\s{}^{\textup t}\bm b&\,\,d\end{pmatrix}.
\end{equation}
In this notation we can express
\begin{subequations}
   \label{list}
	\begin{align}
	   \label{LG_2}
	   \LG &=\left\{Z\in\LGL{n+1}:Z^+ = -Z\right\}
	   \\\label{LP_2}
	   \LP_\s &=\left\{Z\in\LGL{n+1}:Z^+ = -Z \text{ and } Z^- = Z\right\}
	   \\\label{LK}
	   \LK &=\left\{Z\in\LGL{n+1}:Z^+ = -Z \text{ and } Z^- = -Z\right\}
	   \\\label{K}
	   K &=\left\{k\in\LGL{n+1}:k^+k=\1 \text{ and } k^-k=\1\right\}
	\end{align}
\end{subequations}
and we claim that
\begin{equation}
   \label{normalizer_2}
   N(\LG)=\left\{a\in\GL{n+1}:a^+a=\l\1 \text{ for some } \l>0\right\}.
\end{equation}
Indeed, using \eqref{normalizer}, \eqref{LG_2} and the elementary property $(az)^+=z^+a^+$ gets us
\begin{subequations}
   \label{normalizer_computation}
	\begin{align}
	   N(\LG)&=
	   \left\{a\in\GL{n+1}: Z\in\LG \implique aZa\inv\in\LG\right\}
	   \\
	   &=\left\{a\in\GL{n+1}: Z\in\LG \implique (aZa\inv)^+=-aZa\inv\right\}
	   \\
	   &=\left\{a\in\GL{n+1}: Z\in\LG \implique (a^+)\inv Za^+=aZa\inv\right\}
	   \\\label{normalizer_3}
	   &=\left\{a\in\GL{n+1}: Z\in\LG \implique [a^+a,Z]=0\right\}.
	\end{align}
\end{subequations}
Now \eqref{normalizer_3} clearly contains \eqref{normalizer_2}. To see the reverse inclusion we note that if
\begin{equation}
   a^+a=
   \begin{pmatrix}U&\,\,\bm v\\{}^{\textup t}\bm w&\,\,\l\end{pmatrix}
   \quad\text{commutes with every}\quad
   Z=\begin{pmatrix}0&\,\,\bm b\\\s{}^{\textup t}\bm b&\,\,0\end{pmatrix}
   \in\LP_\s
\end{equation}
then
\begin{equation}
   \label{commutator}
   [a^+a,Z]=
   \begin{pmatrix}
      \s\bm v{}^{\textup t}\bm b-\bm b{}^{\textup t}\bm w&\,\,(U-\l)\bm b\\
      \s{}^{\textup t}\bm b(\l-U)&\,\,{}^{\textup t}\bm b(\bm w-\s\bm v)
   \end{pmatrix}
   =0
   \qquad\forall\,\bm b\in\RR^n.
\end{equation}
This gives $U =\l\1$ and $\bm w=\s\bm v$, whereupon \eqref{commutator} becomes the condition that $\bm v{}^{\textup t}\bm b-\bm b{}^{\textup t}\bm v=0$ for all $\bm b\in\RR^n$. As this implies that $\bm v$ is collinear with every $\bm b$ and hence zero, we obtain $a^+a=\l\1$. Moreover it is clear that $\l>0$: if $\s<0$, then $\<x,a^+ax\>_+$ and $\<x,x\>_+$ are simultaneously positive; if $\s>0$ and $\l$ was negative, then $a$ would map $\bigl\{\bigl(\begin{smallmatrix}\bm r\\0\end{smallmatrix}\bigr):\bm r\in\RR^n\bigr\}$ to a $\<\cdot,\cdot\>_+$-positive subspace of dimension $n$, whereas the largest dimension of such a subspace is 1 \eqref{g}. So \eqref{normalizer_2} is proved. 

\subsection{End of proof when $\s\in\RR\smallsetminus\{0\}$ or $\LG=\LK$}

We assume $\s>0$, leaving it to the reader to argue the similar cases $\s<0$ and $\LG=\LK$ which are of little physical interest. We claim that \eqref{normalizer_2} equals
\begin{equation}
   \label{normalizer_4}
   N(\LG)=\left\{a\in\GL{n+1}:a = \sqrt\l k\e{Z} \text{ for some }\l>0, k\in K, Z\in\LP_\s\right\}.
\end{equation}
The theorem follows: indeed, we already know that $K\exp(\LP_\s)\subset G\subset N(\LG)$; and if $a$ in \eqref{normalizer_4} belongs to $G$ then so does $a\e{-Z}k\inv = \sqrt\l\1$, which forces $\l=1$ \eqref{homogeneous_hypothesis}. So there only remains to prove \eqref{normalizer_4}.

To this end we note that if $a$'s expression in \eqref{normalizer_4} holds, then \eqref{list} implies $\smash{\frac1\l a^-a=\e{2Z}}$. 
So for $a$ in \eqref{normalizer_2} we define $Z=\smash{\frac12\log\left(\frac1\l a^-a\right)}$
and $k=\smash{\frac1{\sqrt\l}a\e{-Z}}$
and check:
\begin{compactenum}[(a)]
   \item $Z$ is well-defined: indeed $p=\smash{\frac1\l a^-a}$ is a positive operator on the (positive) inner product space $(\RR^{n+1}, \<\cdot,\cdot\>_-)$, so it has a unique positive logarithm.
   \item $p^+p=\1$: this follows from $a^+a=\l\1$ and $(a^-)^+=(a^+)^-$.
   \item $Z$ is in $\LP_\s$, by \eqref{LP_2}: indeed we have $Z^-=Z$ (positive implies self-adjoint) and $Z^+=\smash{\frac12\log(p^+)}=\smash{\frac12\log(p\inv)}=-\smash{\frac12\log(p)} = -Z$.
   \item $k$ is in $K$, by \eqref{K}: indeed we have $k^+k=\smash{\frac1\l\e{-Z^+}a^+a\e{-Z}}=\smash{\e{Z}\e{-Z}}=\1$ and $k^-k=\smash{\frac1\l\e{-Z^-}a^-a\e{-Z}}=\smash{\e{-Z}\e{2Z}\e{-Z}}=\1$.\qed \end{compactenum}

\begin{remark}
   Once $\l$ is set to $1$, \eqref{normalizer_2} becomes the linear isometry group of the metric $\g_+$ (\ref{metric}, \ref{g}), and \eqref{normalizer_4} gives its well-known Cartan decomposition \cite{Souriau:1970}.
\end{remark}

\begin{remark}
   If we replace the right-hand side of \eqref{homogeneous_hypothesis} by $\smash{\left(\begin{smallmatrix}\SO{n}&0\\0&1\end{smallmatrix}\right)}$, then considering the matrix
	\begin{equation}
	   \exp\begin{pmatrix}0&\,\,\bm b\\\s{}^{\textup t}\bm b&\,\,0\end{pmatrix}
	   =\begin{pmatrix}
	      \1 - \bm u{}^{\textup t}\bm u + \cos(\|\bm b\|/\C)\bm u{}^{\textup t}\bm u&\,\,\,\sin(\|\bm b\|/\C)\C\bm u\\
	      \hspace{3.7em}-\sin(\|\bm b\|/\C){}^{\textup t}\bm u/\C&\,\,\,\cos(\|\bm b\|/\C)
	   \end{pmatrix}
	\end{equation}
   $(\bm b=\|\bm b\|\bm u)$ for $\|\bm b\|/\C\in(2\ZZ+1)\pi$ shows that the case $\s=-1/\C^{2}<0$ must be suppressed from the conclusion of Theorem \ref{homogeneous_theorem}. If we further strengthen the hypothesis by requiring
   \begin{equation}
      \label{Gorinis_hypothesis}
   	G\cap\begin{pmatrix}\GL{n}&\,0\\{}^{\textup t}\RR^n&\,\RR^\times\end{pmatrix}
   	=\begin{pmatrix}\SO{n}&\,\,0\\0&\,\,1\end{pmatrix},
   \end{equation}
   then the case $\s=\infty$ must also be suppressed from the conclusion. This is the actual formulation of Gorini \cite[Thm 1]{Gorini:1971}, \cite{Gorini:1973}.
\end{remark}

\begin{acknowledgements}
   We wish to thank Arnaud Beauville and Antoine Derighetti for very helpful indications.
\end{acknowledgements}

\end{document}